\documentstyle[epsfig]{aipproc}

\begin{document}
\title{Choosing a measure of GRB brightness\\ 
that approaches a standard candle
}

\author{J-L. Atteia$^*$
}
\address{$^*$Centre d'Etude Spatiale des Rayonnements (CNRS/UPS)\\
BP 4346, F 31028 Toulouse Cedex 4, France\\
}

\maketitle

\begin{abstract}
Studies using the GRB brightness as a distance indicator 
require a measure of brightness with a small intrinsic 
dispersion (close to a standard candle). 
There is unfortunately no general agreement on the definition 
of such a quantity. 

We show here that the comparison of the size-frequency curves 
obtained with various measures of brightness can be used 
to select the quantity which is closer to a standard candle.
Our method relies on a few general assumptions 
on the burster spatial distribution, 
namely that nearby bursters are homogeneously distributed 
in an Euclidean space with no density or luminosity evolution.
We apply it to 5 measures of GRB brightness in the Current 
BATSE Catalog and we find that the GRB size-frequency distribution
depends significantly on the energy window used to measure the GRB brightness.
The influence of the time window being, in comparison, negligible. 
Our method suggests that the best distance indicator in this Catalog 
is the fluence measured below 100 keV, indicating that GRB luminosities
have a smaller {\it intrinsic} dispersion below 100 keV than above. 
\end{abstract}

\section*{Principle of the method}

We start from the basic observation that GRB size-frequency curves
(log(N)-log(flux)) represent a convolution of the burster distributions
in distance and luminosity. 
These two distributions, however, cannot be easily disentangled and
if we want to learn something about the GRB luminosity function,
we have to make assumptions on the burster radial distribution.

Practically we assume that :

\begin{itemize}
\item The slope -3/2 observed at the bright end of the size-frequency
distribution reflects an actual spatial homogeneity of nearby 
bursters (in an Euclidean space). 

\item BATSE is sensitive enough to detect all the bursts emitted inside 
this homogeneous region (we show below that this second assumption is
fully justified).
\end{itemize}

At this point we should stress that the assumptions stated above
define the validity of our method.
In particular our method is not relevant if the intensity distribution 
of bright bursts is significantly affected by source evolution\cite{wij97}.
Under these assumptions, the GRB size-frequency curve follows 
the law expected for spatially homogeneous sources (with a slope~=~-3/2) 
at its bright end, and progressively deviates from this law for 
fainter bursts.
We demonstrate below that the number of bursts seen in the homogeneous
part of the size-frequency curve depends on the intrinsic width 
of the burster's luminosity distribution, and we explain how this number 
can be used to compare the intrinsic dispersion of various measures 
of the GRB luminosity.
\medskip
Lets $N_h$ be the total number of sources in the homogeneous region 
that have emitted a burst and $n_{3/2}$ the number of sources
that we measure in the homogeneous part of the size-frequency curve.

\begin{itemize}
\item For {\it standard candles}, the size-frequency curve deviates 
from homogeneity only when all the sources in the homogeneous region 
have been sampled, and we have $\bf n_{3/2} = N_h$.
\item For sources with a {\it broad luminosity dispersion}, 
the size-frequency curve deviates from homogeneity when the 
(intrinsically) bright sources become detectable beyond the 
homogeneous region.
If the luminosity dispersion is broad, this occurs while only part 
of the sources in the homogeneous region are detectable, and we 
expect $\bf n_{3/2} < N_h$.
\end{itemize}

\begin{figure}
\centerline{\epsfig{file=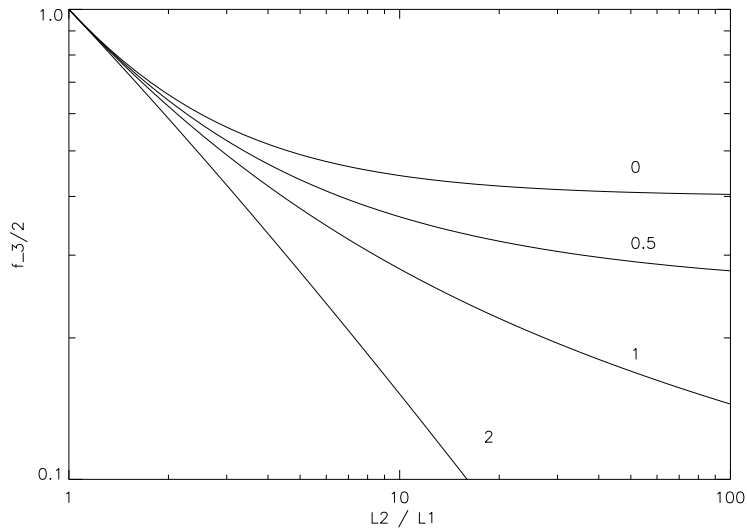,height=7.5cm,width=10.5cm}}
     \caption{Influence of an intrinsic luminosity distribution on the number
of GRBs seen in the homogeneous part of the size-frequency curve.
We assume that the luminosity function has the form 
$n(L) \propto L^{- \alpha}$ between $L_1$ and $L_2$.
The figure displays $f_{3/2}$ as a function of $L_2 / L_1$, where
$f_{3/2}$ is the ratio of the number of GRBs seen in the homogeneous
part of the size-frequency curve to the number expected for 
standard candles. Note that $f_{3/2}$ decreases when $L_2 / L_1$ 
increases (the curves are for $\alpha = 0, 0.5, 1, 2$, top to bottom).}
         \label{rapn}
\end{figure}

This situation is illustrated on Fig.~1, which displays 
$ f_{3/2} = n_{3/2}~/~N_h$ as a function of the  
luminosity dispersion of the sources.
This figure shows that the ratio $f_{3/2}$ decreases 
as the luminosity dispersion increases.

In the real world we do not know $N_h$, so we cannot decide how close
to a standard candle is a given measure of the GRB luminosity.
We can, however, compare the values of $n_{3/2}$ obtained for different 
definitions of the GRB luminosity, in order to decide which one
has the smallest intrinsic dispersion.

\section*{\bf Application to the Current BATSE 
Catalog\footnote{A\lowercase{s available on 1997, June 30th}}}

Figure~2 illustrates the application of this method to the bursts in
the Current BATSE GRB Catalog \cite{mee97}.
We have chosen 5 measures of the GRB luminosity given in this Catalog :

- P64, the peak flux in the 50-300~keV energy range accumulated over 64~ms. 

- P1, the peak flux in the 50-300~keV energy range accumulated over 1~s. 

- F12, the fluence in the energy range 25-100~keV.

- F23, the fluence in the energy range 50-300~keV.

- F34, the fluence above 100~keV.

\medskip
Figure~2 displays the size-frequency distribution of 1103 GRBs 
with T90~$>$~2~s, the extrapolation of the homogeneous part of the curve,
and an estimate of $n_{3/2}$.

\medskip
{\bf Influence of the time window.} 
In the three panels of the first row (P64, P1 and F23), the luminosity 
is measured in the same energy domain (50-300~keV) but with different 
integration times (64~ms, 1~s and the entire burst duration). 
The deviation from homogeneity occurs at P64~=~6.9~ph~cm$^{-2}$~s$^{-1}$,
P1~=~6.6~ph~cm$^{-2}$~s$^{-1}$ and F23~=~10$^{-5}$~erg~cm$^{-2}$.
This justifies a-posteriori our second assumption, since the catalog
of BATSE is over 99\% complete at and above 
these intensities.
The values of $N_b$ and their 90\% error bars are respectively
{\bf 129} [78-151], {\bf 95} [74-130] and {\bf 139} [100-210], 
for P64, P1 and F23
(the error bars are computed from a bootstrap resampling of the 
size-frequency curve).
These three measures of intensity show little difference
from the point of view of our analysis, suggesting
that the choice of the time window is not  
crucial in the search for a standard candle luminosity.

\medskip
{\bf Influence of the energy window.} 
In the three panels of the second row (F12, F23 and F34), 
the luminosity is integrated over the entire burst  
but in different energy ranges (25-100, 50-300 and $>$100~keV).
The departure from homogeneity occurs at F12~=~2.6~10$^{-6}$~erg~cm$^{-2}$,
F23~=~10$^{-5}$~erg~cm$^{-2}$ and F34~=~3.9~10$^{-5}$~erg~cm$^{-2}$.
We checked that the trigger efficiency of BATSE for GRBs 
in the homogeneous part of the size-frequency is greater than 98\% 
in the 3 cases.
The values of $N_b$ and their 90\% errors are respectively
{\bf 254} [190-318], {\bf 139} [100-210] and {\bf 76} [60-112].
These numbers show that the GRB size-frequency distribution contains
about 3 times more GRBs in its homogeneous part 
when F12 instead of F34 is used to measure the burst intensity.
This suggests that the low energy fluence (below
100 keV) is significantly closer to a standard candle 
than the fluence measured at higher energies.\footnote{Our method 
provides almost identical values of $n_{3/2}$ (237 and 224) for
the fluence distribution in 25-50~keV and 50-100~keV.}

Contrary to the time window, the energy window appears as a 
crucial parameter in the search for a luminosity with a small
intrinsic dispersion. 

\section*{Discussion} 

We have proposed a new method to compare the intrinsic dispersion 
of various measures of GRB brightness.
An essential assumption of this method is that the slope -3/2 
at the bright end of the size-frequency curve is due to 
the spatial homogeneity of nearby bursters.
In the rest of the discussion we assume that source evolution does not 
dominate the size-frequency distribution of bright GRBs.

The fact that the width of the $\gamma$-ray burst luminosity function
changes with the energy calls for a careful definition of the GRB luminosity
when the intensity is taken to reflect the distance to the sources 
(e.g. to compare the properties of nearby and distant bursters). 
Measures at low energies (below 100~keV) are clearly preferred.

We also note that the combination of {\it i)} a broad luminosity function 
and {\it ii)} a spatial density which varies with the distance 
has interesting consequences for the comparison of faint and bright GRBs.
Intrinsically bright bursts are detected to large distances 
(typically larger than the size of the homogeneous region) where
the burster spatial density decreases rapidly.
Intrinsically faint bursts on the other hand are only visible 
to much smaller distances where the burster density is constant
(if we remain in the homogeneous region) or slowly decreasing.
As a consequence, going to lower intensities increases the number
of bright GRBs much less (in percentage) than the number of 
intrinsically faint bursts. 
In other words, burst classes based on the {\it observed} brightness 
do not contain the same proportion of {\it intrinsically} bright bursters.
This changing proportion may produce brightness-dependent 
burst properties (spectral and/or temporal) 
which could strengthen or counteract cosmological effects.
If, as suggested by this study, the GRB luminosity function is broader
at high energies, we predict that average burst properties
will appear {\it more brightness-dependent} when brightness is measured 
at higher energies. 

\medskip
Finally, while we do not address here the reasons which make 
the luminosity at low energies a better standard candle,
we note that this behaviour could well be explained by 
an anisotropy of the emission above $\sim$100~keV.
Such an anisotropy would make the flux at high energies 
dependent on the aspect of the source.
From the point of view of the size-frequency distribution, 
a beaming of the emission is equivalent to the existence
of a luminosity function. 
Hence a beaming factor which changes with the energy 
may just appear as an energy dependent luminosity function, 
in accordance with what we observe.

{\it Acknowledgements}

The author is grateful to the BATSE team for making the BATSE data 
rapidly available through the Current BATSE Catalog and to J-P. Dezalay 
for helpful discussions on the interpretation of size-frequency distributions.

 \begin{figure}
    \epsfig{file=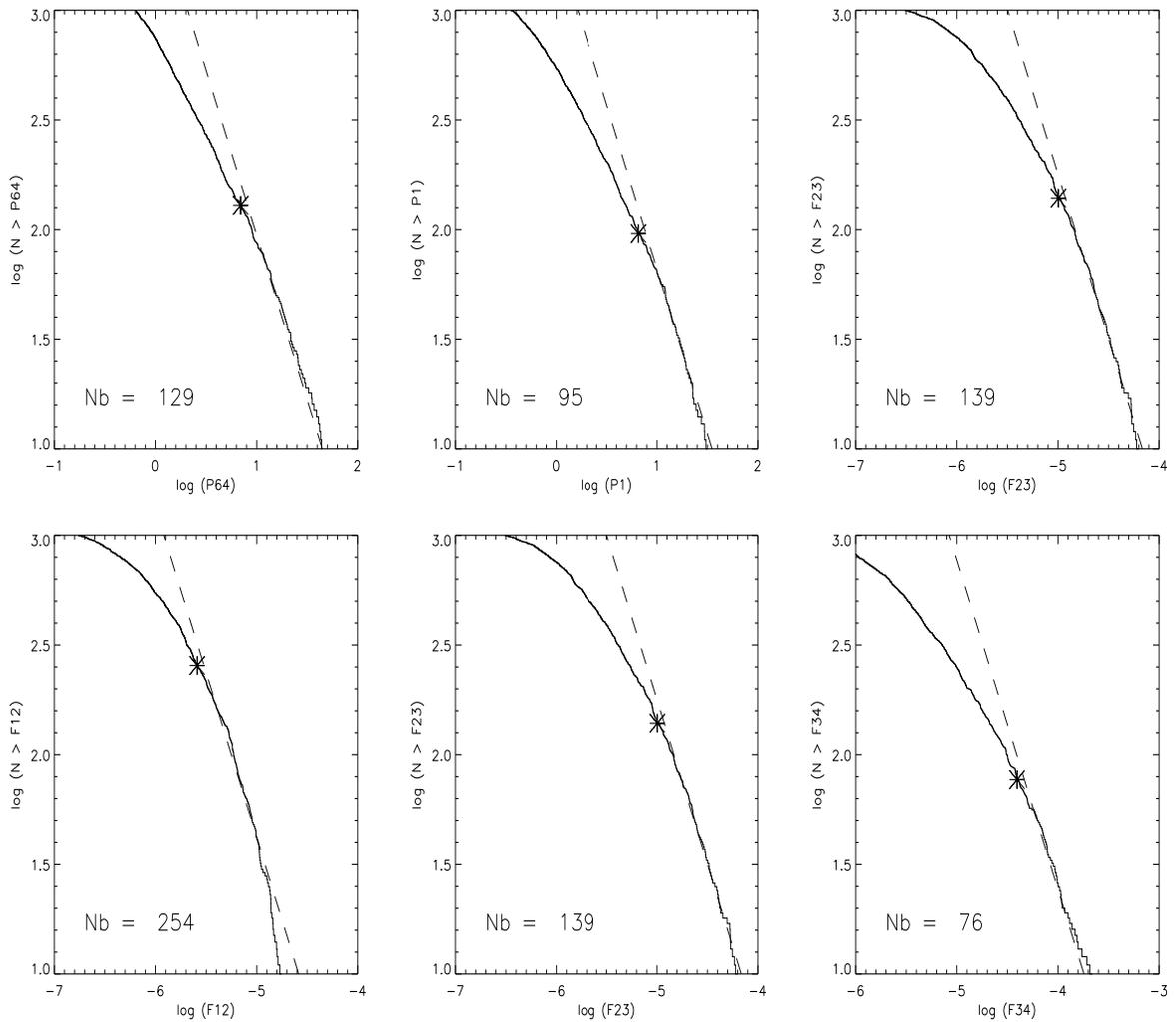,height=14cm,width=16cm}
     \caption{Value of $n_{3/2}$ for various measures of the GRB brightness (see text).}
         \label{Figsix}
    \end{figure}

\end{document}